\documentstyle[emulateapj,psfig,apjfonts]{article}
\def\etal{et al.\,}

\def\be{\begin{equation}}
\def\ee{\end{equation}}

\def\refnew#1{(\ref{#1})}

\begin{document}

\title{\mbox Inferring Initial Spin Periods for Neutron Stars in Composite
Remnants}

\lefthead{Inferring neutron star initial spin periods}
\righthead{van der Swaluw \& Wu}

\author{E. van der Swaluw \altaffilmark{1}, Y. Wu \altaffilmark{2}}
 
\altaffiltext{1}{E.vanderswaluw@astro.uu.nl, Astronomical Institute, Utrecht University, P.O. Box
80000, 3508 TA Utrecht, the Netherlands}

\altaffiltext{2}{wu@cita.utoronto.ca, Canadian Institute for Theoretical Astrophysics,
        University of Toronto, 60 St. George Street, Toronto, Ontario
        M5S 3H8, Canada}

%\date{Received..........; accepted..........}
%\maketitle 

%\clearpage

\begin{abstract}
We propose a method to infer the initial spin period of pulsars
residing in composite supernova remnants. Such a remnant consists of
both a plerionic and a shell type component, corresponding
respectively to the pulsar wind nebula driven by the spindown
luminosity of the central pulsar, and the blastwave bounding the
supernova remnant.  Theoretical investigations including
hydrodynamical simulations have shown that at late times ($\sim$ 1,000
- 10,000 years), a simple scaling law connects the radius of the
supernova shell to the radius of the plerion. The energy content of
the plerion and the total mechanical energy of the supernova remnant
enter into this scaling law. One can use this scaling law to estimate
the initial spin period of pulsars residing in composite remnants.  We
discuss potential pitfalls of this method, including the effect of a
small remnant age and of strong radiative losses in the plerion.
\end {abstract}
%\clearpage

\section{Introduction}

Is a typical neutron star born spinning close to break-up ($P_0 \sim
0.8$ ms, Haensel \etal \cite{1995A&A...296..745H}). The answer to this
question reveals much about the angular momentum evolution of neutron
star progenitors (Heger \etal \cite{2000ApJ...528..368H}), as well as
about the processes that occur during the supernova explosion (Spruit
\& Phinney \cite{1998Natur.393..139S}, Lai \& Goldreich
\cite{2000ApJ...535..402L}). If the initial spin-rate is close to the
break-up rate, the nascent neutron star may undergo a Rossby-wave
instability (Lindblom, Owen \& Morsink \cite{1998PhRvL..80.4843L}) and
emit gravitational radiation. Moreover, a rapidly rotating neutron
star can obtain a substantial kick velocity through the
electromagnetic rocket effect (Harrison \& Tademaru
\cite{1975ApJ...201..447H}). This can provide a natural explanation
for the apparent alignment between the spin axis and pulsar proper
motion (Lai \etal \cite{2001ApJ...549..1111}) in the Crab (Caraveo \&
Mignani \cite{1999AA...344..367C}) and Vela pulsar (Pavlov \etal
\cite{2000AAS...196.3704P}).

A number of studies have been dedicated to the determination of
initial spin-rates. Those few young pulsars which have measured
braking indices and estimated ages seem to possess initial spin
periods from $19$ ms (Crab, Lyne \etal \cite{1993MNRAS.265.1003L}) to
$63$ ms (PSR B1509-58, Kaspi \etal
\cite{1994ApJ...422L..83K}).\footnote{The LMC x-ray pulsar J0537-6910
($16$ ms, Marshall \etal \cite{1998ApJ...499L.179M}) should be born at
shorter than $10$ ms for reasonable values of the braking index.}
Since only rapidly rotating pulsars have measurable braking indicies,
these values may not be representative for the general population. For
these, Phinney \& Blandford (\cite{1981MNRAS.194..137P}) and
Vivekanand \& Narayan (\cite{1981JApA....2..315V}) have developed the
pulsar-current analysis. Unfortunately, this method is subject to
small-number statistics (Lorimer \etal
\cite{1993MNRAS.263..403L}). Analysis based on pulsar luminosity
functions (Emmering \& Chevalier \cite{1989ApJ...345..931E}, Narayan
\cite{1987ApJ...319..162N}) raised the possibility of `injection': the
majority of the pulsars may be born with periods as slow as several
hundred milliseconds. This hypothesis is corroborated by the apparent
paucity of plerions around neutron stars (Srinivasan \etal
\cite{1984JApA....5..403S}). A plerion derives energy from the
rotational energy of the central pulsar. The overall conclusion from
pulsar studies seems to favour a slow initial spin rate.

In this article, we propose a new method for inferring the initial
spin-rates of some neutron stars -- the small population that reside
within composite supernova remnants.  A composite supernova remnant
(SNR) includes both a plerionic (filled-centered) component (Weiler \&
Panagia \cite{Weiler}) and a shell component. The first corresponds to
the pulsar wind nebula (PWN), while the latter corresponds to the
blastwave of the supernova remnant (SNR) propagating through the
interstellar medium (ISM). They are observationally distinguished by a
different spectral power-law index at radio frequencies. The dynamics
of such a composite system has been extensively studied, e.g. Reynolds
\& Chevalier (1984); Chevalier \& Fransson (1992).

Recently, van der Swaluw \etal (\cite{eric}) presented hydrodynamical
simulations and analytical arguments which explicitly relate the
radius of the PWN to that of the SNR. These authors find that the two
radii are roughly proportional some time after the initial explosion,
with the proportionality constant determined by the ratio of the
pulsar spin-down energy and the total mechanical energy of the
supernova event.  This relation forms the theoretical basis for the
current paper.  In this article, we follow van der Swaluw \etal
(\cite{eric}) in restricting ourselves to spherically symmetric
systems.

\section{Evolution of a PWN in a SNR}

As a pulsar spins down, its rotational energy is largely
deposited into the surrounding medium, driving a relativistic pulsar 
wind into the tenuous bubble of stellar debris left behind by the supernova 
blastwave.  We approximate the spin-down luminosity of the pulsar wind by 
that of a rotating magnetic dipole:
\begin{equation}
	L(t) = {L_0}/\left(1+{t/\tau}\right)^2.
\label{pulsarlum}
\end{equation}
The total spin-down energy is $E_{\rm sd} = L_0 \tau$. We neglect the
dynamical influence of the pulsar wind on the SNR, assuming that
$E_{\rm sd}$ is well below the total mechanical energy of the
supernova explosion, $E_0 \sim 10^{51}$ erg.

The evolution of a PWN inside a SNR can be divided into two stages
(Reynolds \& Chevalier 1984; Chevalier \& Fransson 1992).  In the
first stage, the PWN expands supersonically within the bubble blown
out by the SNR. The second stage commences at $t \equiv t_{\rm ST}$
when the PWN encounters the reverse shock of the SNR. The reverse
shock heats the interior of the SNR and causes the expansion of the
PWN to become subsonic.  In this stage the expansion of the PWN is
regulated by the expansion of the SNR, which is described by a
self-similar Sedov-Taylor solution.  In the following we summarize
some relevant results from Chevalier \& Fransson (1992) and van der
Swaluw \etal (\cite{eric}).

\subsection{Analytical Relations
\label{subsec:analytical}}

We focus on the subsonic expansion stage of the PWN which occurs when
the SNR has relaxed to the Sedov solution. We assume that the pulsar
wind has deposited most of its energy into the PWN ($t \gg \tau$).  We
find that the radius of the PWN ($R_{\rm pwn}$) scales roughly
linearly with the radius of the SNR ($R_{\rm snr}$). This scaling
arises from the condition of pressure equillibrium between the PWN
interior and the interior of the SNR.  We approximate the pressure
which confines the PWN by the central pressure of the SNR, using the
Sedov solution (Sedov \cite{sedov1958}),
\begin{equation}
	P_{\rm snr}\simeq 0.074 E_0/R_{\rm snr}^3 \; , 
\label{eq:psnr}
\end{equation}
where $E_0$ is the total mechanical energy of the SNR. Inside the PWN,
pressure is quickly equilibrated due to the high sound speed $\sim
c/\sqrt{3}$ of the relativistic fluid. The interior pressure is
related to the energy content of the PWN, $E_\ast$, by
\begin{equation}
	P_{\rm pwn}\simeq {3({\gamma -1})\over 4\pi} {E_\ast\over
	R_{\rm pwn}^3},
\label{eq:PWNev}
\end{equation}
where $\gamma$ is the adiabatic index. Let $E_{\rm pwn}$ be the total
amount of energy injected into the PWN by the central pulsar. Part of
this energy is used to perform work on the surrounding medium as the
PWN expands. Despite some compression of the PWN by the reverse shock
around $t= t_{\rm ST}$, $E_\ast$ falls below $E_{\rm pwn}$ (see Fig. 9
of van der Swaluw \etal (\cite{eric})).  We therefore write $E_\ast
=\eta_2 E_{\rm pwn}$ with $\eta_2 < 1$ decreasing over time. Imposing
pressure equilibrium, we find
\begin{equation}
	R_{\rm pwn} = \bar C \: \left( {E_\ast/E_0} \right)^{1/3}
	\: R_{\rm snr} \; ,
\label{eq:PWNra}
\end{equation}
where $\bar C\simeq 1.02$ for a relativistic fluid ($\gamma = 4/3$)
and $\bar C\simeq 1.29$ for a non-relativistic fluid ($\gamma = 5/3$).

In the absence of radiative losses, the subsonic expansion of the PWN
is adiabatic, $P_{\rm pwn} V_{\rm pwn}^{\gamma}$ = constant, where
$V_{\rm pwn} \propto R_{\rm pwn}^{3}$ is the volume of the PWN.
Combining this relation with equation \refnew{eq:PWNev}, we find that
$E_\ast$ falls off with time as $t^{-3/10}$ for a relativistic fluid.
This yields an expansion law for the radius of the PWN, $R_{\rm
pwn}\propto t^{3/10}$, whereas the radius of the SNR scales as in a
Sedov solution ($R_{\rm snr}\propto t^{2/5}$). So, roughly, $R_{\rm
pwn} \propto R_{\rm snr}$.

We introduce two additional dimensionless parameters, $\eta_1$ and
$\eta_3$, and rewrite equation \refnew{eq:PWNra} into a form that can
be more easily compared with simulations,
\begin{equation}
	R_{\rm pwn}(t) = \eta_3(t) \: \left({\eta_1 E_{\rm
	sd}}/{E_0}\right)^{1/3}\: R_{\rm snr}(t).
\label{eq:introduce}
\end{equation}
The parameter $\eta_1$ is a constant and relates the total energy
input into the PWN, $E_{\rm pwn}$, to the total spin-down energy of
the pulsar, $E_{\rm sd}$ as $E_{\rm pwn} = \eta_1 E_{\rm sd}$.  This
takes into account possible radiative losses in the PWN, and other
inefficiencies in the conversion of the spin-down energy into
mechanical energy.  For instance, a fraction of the neutron star
spin-down energy may escape directly from the pulsar as high energy
radiation. We set $\eta_1 = 1$ in this article, but discuss the case
when it is much less than unity. The second parameter is defined as
$\eta_3(t) =\bar C \eta_2^{1/3} \leq \bar C$.  The maximum value of
$\eta_3 = \bar{C}$ is used when we determine the initial spin
periods of pulsars driving composite remnants.

%Equation \refnew{eq:introduce} can be applied formally to the early
%(supersonic) stage of the PWN evolution. But the interpretations for
%$\eta_1$ and $\eta_3$ become unclear.  In the following, we
%demonstrate the behaviour of $\eta_3$ throughout the PWN evolution
%using results from hydrodynamical simulations.
%wyq: changed the above paragraph
The derivation leading up to equation \refnew{eq:introduce} is invalid
during the earlier supersonic evolution. We retain equation
\refnew{eq:introduce} with the sole purpose of defining $\eta_3$
during these stages. We proceed to demonstrate the behaviour of
$\eta_3$ throughout the PWN evolution with results from hydrodynamical
simulations.

\subsection{Hydrodynamical Simulations
\label{subsec:simulation}}

Van der Swaluw \etal (\cite{eric}) have presented hydrodynamical
simulations of a pulsar wind nebula inside a supernova remnant.  In
these simulations, the gasdynamical equations are integrated in a
spherically symmetric configuration, using the Versatile Advection
Code (VAC) developed by G\'abor T\'oth at the Astronomical Institute
Utrecht (T\'oth and Odstr\v cil \cite{Toth2}). Lacking the
possibility of treating a relativistic fluid, the pulsar wind has been
implemented as a cold non-relativistic wind ($\gamma = 5/3$) with a
terminal velocity equal to the speed of light, $v_\infty = [2L(t)/\dot
M_{\rm pw}(t)]^{1/2} \simeq c$, where $\dot M_{\rm pw}(t)$ is the mass
ejection rate into the pulsar wind bubble, and $L(t)$ is the pulsar
spin-down luminosity as given by equation \refnew{pulsarlum}.
	
\vbox{
\begin{center}
\leavevmode \hbox{ \epsfxsize=0.85\hsize \epsffile[80 380 570
700]{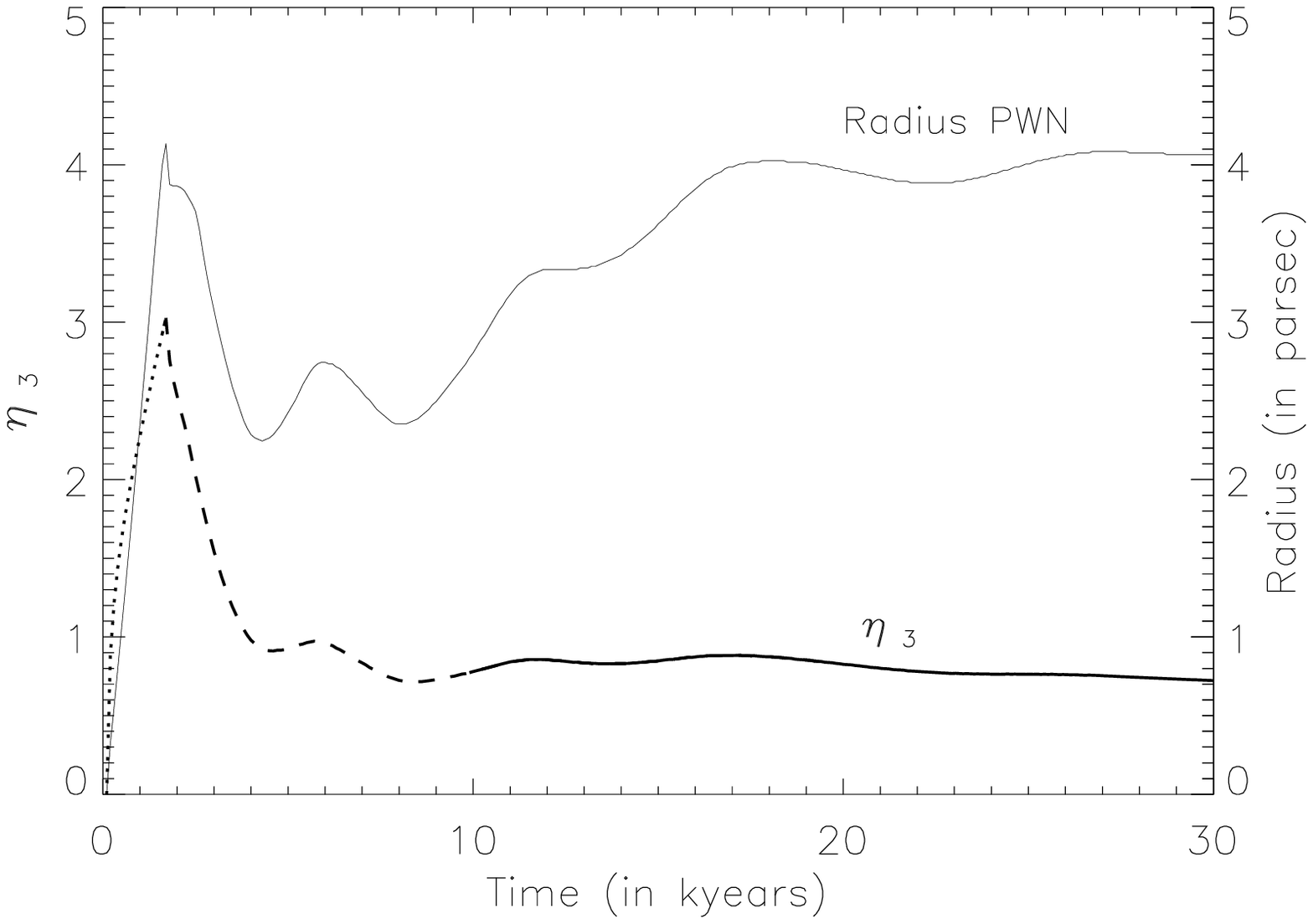}} \figcaption{\footnotesize Results of numerical
simulation showing the time evolution of the pulsar wind nebula.
Plotted here are its radius (in units of parsec, thin continued line)
and the dimensionless parameter $\eta_3$ (thick line). The latter is
disected into three parts: a supersonic expansion stage when the PWN
is bounded by a shock (dotted section), in unsteady transition when
undergoing reverberation with the supernova reverse shock (dashed
section), and subsonic expansion (solid section). The parameters
adopted in this simulation is: maximum spin-down luminosity $L_0 =
5\times 10^{38}$ ergs/s, spin-down time $\tau=600$ yrs, supernova
explosion energy of $E_0 = 10^{51}$ ergs and ISM density of $10^{-24}$
g/${\rm cm}^3$. We find $t_{\rm ST} \sim 10^3 {\rm yrs}$ in this
case. Note that $\eta_3$ is not physical in the early expansion stages
but is only formally defined by equation \refnew{eq:introduce}.  For
details, see van der Swaluw \etal (\cite{eric}).
\label{fig:numerical}}
\end{center}}

Figure \ref{fig:numerical} presents relevant results from one of these
simulations. The numerical parameters adopted in this simulation are
listed in the caption.  Taking different parameters will not
qualitatively change the overall behaviour of the system, but it will
affect the early (supersonic) evolution and the moment when subsonic
expansion commences ($t_{\rm ST}$).  For example, $t_{\rm ST}$ is
larger if the supernova explosion is more energetic, if the supernova
ejecta mass is larger, or if the interstellar medium density is
smaller ($t_{\rm ST} \propto E_{0}^{-1/2} M_{\rm ej}^{5/6} \rho_{\rm
ism}^{-1/3}$, e.g. McKee \& Truelove (1995)).  Since this is a
non-relativistic simulation ($\gamma = 5/3$), we expect $\eta_3 \leq
\bar C = 1.29$ in the subsonic expansion stage. This is indeed
observed. Therefore in the following investigation we assume a maximum
value for $\eta_3$ of $1.02$, as is appropriate for the subsonic
expansion stage of a PWN in a SNR with $\gamma =4/3$.

\section{Inferring Initial Spin Rates
\label{sec:result}}

Writing the spin-down energy as $E_{\rm sd} = (\Omega_0^2 -
\Omega_t^2) I/2$, we obtain the following expression for the pulsar's
initial spin period,
\begin{equation} 
	P_0 = 2\pi\left({2E_0\over {\eta_1 I}} \left( {R_{\rm
	pwn}\over\eta_3 R_{\rm snr}}\right)^3 +
	\left({{2\pi}\over{P_t}}\right)^2 \right)^{-1/2},
\label{eq:includept}
\end{equation}
where the spin-periods are $P_0 \equiv 2 \pi/\Omega_0$, $P_{\rm t}
\equiv 2 \pi/\Omega_t$ and $I$ is the moment of inertia of the neutron
star.  The radius of a PWN relative to its associated SNR shell can be
used to infer the initial spin-period of the pulsar, assuming
$\eta_1$, $\eta_3$, $E_0$ and $\Omega_t$ are known or can be
estimated.

\vbox{
\begin{center}
\leavevmode \hbox{ \epsfxsize=0.80\hsize \epsffile[20 195 570
670]{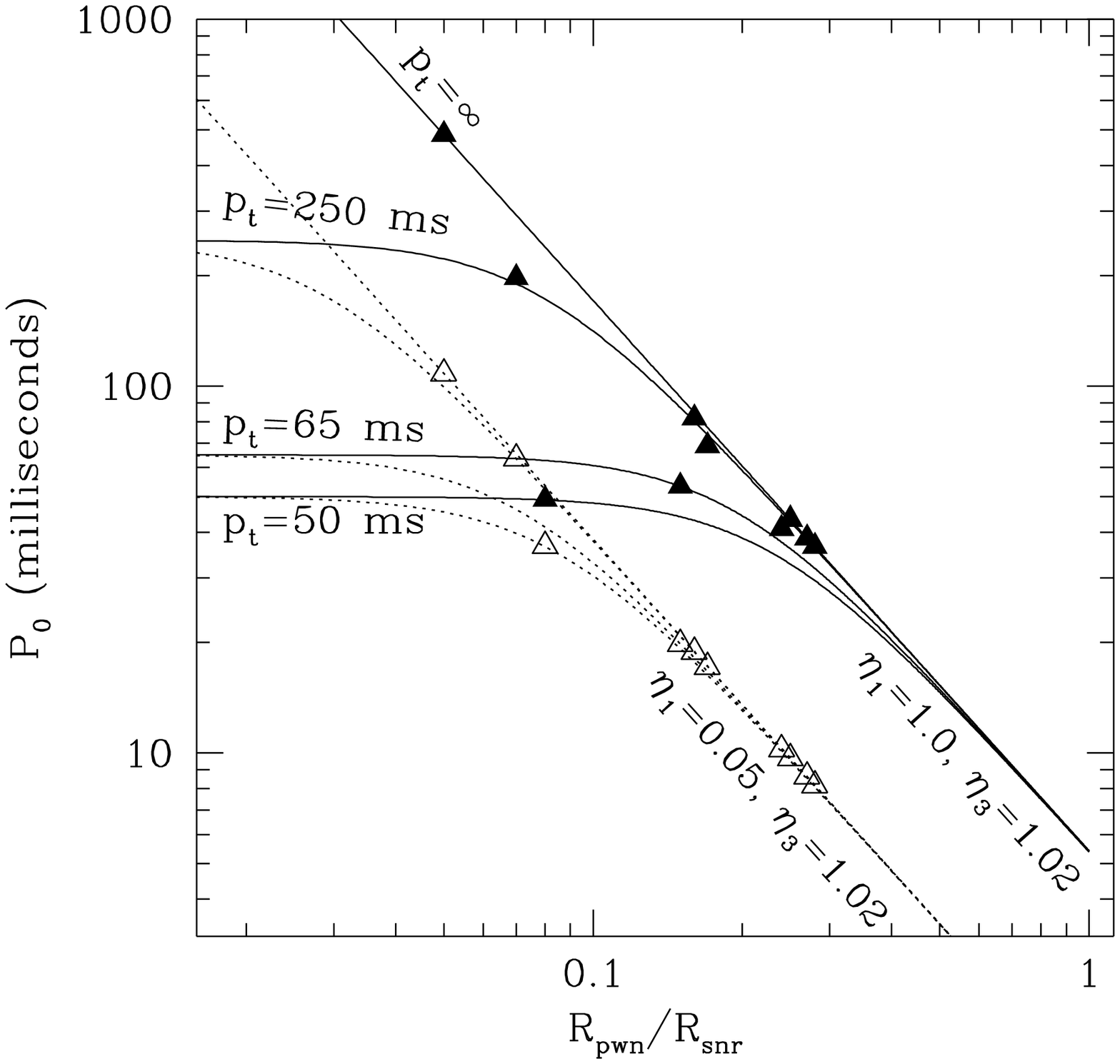}} \figcaption{\footnotesize Initial spin periods inferred
from the ratio $R_{\rm pwn}/R_{\rm snr}$ using equation
\refnew{eq:includept}.  When assuming no radiative loss ($\eta_1 =
1$), we obtain solutions as depicted by the family of solid lines,
with different lines applicable for different current spin periods,
$P_t$. The family of dashed lines are for $\eta_1 = 0.05$. Filled and
open triangles represent systems listed in Table \ref{table:table2}
for $\eta_1 = 1$ and $\eta_1 = 0.05$, respectively.  When $P_0 \leq 4$
ms, the initial rotational energy of the puslar exceeds the total
mechanical energy in the SNR itself. Equation \refnew{eq:includept}
fails and the SNR may be blown away by the PWN. None of the systems we
examined lies close to this limit.
\label{fig:result}}
\end{center}}

The radius of PWN should be taken from radio measurements.
Synchrotron-radiating electrons in PWN have a typical life-time
(Rybicki \& Lightman \cite{rybicki}):
\begin{equation}
\tau_{{\rm 1/2}} = {6\pi m_{\rm e}c\over \sigma_{\rm T}B^2\Gamma} \sim
                              10^5 B_{m {\rm G}}^{-3/2} \nu_{\rm
                              MHz}^{-1/2}\,\,\, {\rm yrs},
\label{eq:lifetime}
\end{equation}
where $\sigma_{\rm T}$ is the Thomson cross-section, $\Gamma$ is the
electron Lorentz factor, $\nu$ is the typical synchrotron frequency
scaled at the radio wavelength, and $B$ is the magnetic field strength
scaled by a value appropriate for plerions. Such a long life-time
implies that radio observations `see' electrons that were accelerated
during the early energetic stage when the pulsar deposits most of its
rotational energy into the pulsar wind. They therefore delineates the
true spatial extent of the plerion. In contrast, X-ray electrons loose
their energy quickly. The X-ray plerion is produced by electrons
recently accelerated near the pulsar, so one ecpects the plerion to be
smaller in X-rays.
%more so for older systems.
\begin{table*}
\begin{center}
{
\small
\begin{tabular}{|l|l|c|c|c|c|}
\hline\hline
Supernova & \multicolumn{2}{c} {Plerion} & \multicolumn{2}{c} {Pulsar} & age\\

Designation & $R_{\rm pwn}/R_{\rm snr}$ & $P_0$ (in msec) & $P_{t}$ (in msec) &
$P_0^\prime$ (in msec) & (kyr)\\ 
\hline
G$0.9+0.1$ & $0.25$(Helfand \& Becker \cite{1987ApJ...314..203H}) & $43$
 & --  & -- & -- \\
%checked
%G$6.4-0.1$ &     &    &     \\ %W28, associated with a pulsar?
% 40' diameter radio shell, 5'' flat component (thermal?), Andrews \etal 1983
G$16.7+0.1$ & $0.25$(Helfand \etal\cite{1989ApJ...341..151H}) & $43$&
-- & -- & -- \\
%checked
G$29.7-0.3$ & $0.16$(Becker \& Helfand \cite{1984ApJ...283..154B}) &
$82$ & $325$[PSR 1846-0258] & -- & $0.7$ \\ % Kes 75
%checked, Gottelf etal 2000 gives pulsar detection, and B=5x10^{13}G
G$34.7-0.4$ & $0.07$(Frail \etal \cite{1996ApJ...464L.165F}) & $197$
& $267$[PSR 1853+01]  & -- & -- \\ % W44,3C392
%checked, 30' shell, 2'- 3' center, no mention of spin-down behaviour
%G68.9+2.8 & $0.14$ & $39.5$[PSR1951+32] & --  &  ??  \\ % G69.0+2.7 CTB 80?
%the structure is irregular, with central/surrounding 0.5/8/40
%G$291.0-0.1$ & $0.41$  & --   & $17.4$   & ??   \\ %MSH 11-62
% unclear
%Roger \etal (MNRAS1986.219.815) claimed a centrally filled source of 9'x4'
% but x-ray (Harrus \etal 1998ApJ499,273) has thermal size 6' and synchrotron 3'
% (resolution of ASCA), Whiteoak & Green unclear radio map
G$293.8+0.6$ & $0.25$(Whiteoak \& Green \cite{1996AAS..118..329W}) &
$43$ & -- & -- & -- \\
%checked, Fig. 3 of the reference
G$322.5-0.1$ & $0.27$(Whiteoak \cite{1992MNRAS.256..121W}) & $39$ & --
& -- & -- \\
%checked
G$326.3-1.8$ & $0.28$(Kassim \etal \cite{1993ApJ...419..733K}) & $37$
& -- & -- & -- \\ % MSH 15-56
%checked
G$327.1-1.1$ & $0.25$(Sun \etal \cite{1999ApJ...511..274S}) & $43$ &
-- & -- & -- \\
%checked
G$351.2+0.1$ & $0.05$(Becker \& Helfand \cite{1988AJ.....95..883B}) &
$484$ & -- & -- & -- \\
%checked
\hline
G$263.9-3.3$ & $0.24$(Milne \cite{1980AA....81..293M}) & $41$ &
$89$[Vela] & $52$(Lyne \etal \cite{1996Natur.381..497L}) & $11.3$ \\
%checked %p_0=52?
G$11.2-0.3^\ast$ & $0.15$(Vasisht \etal \cite{1996ApJ...456L..59V}) &
$53$ & $65$ & $63$(Torii \etal \cite{1999ApJ...523L..69T}) & $1.6$\\
%p_0 = 62.8 ms
%checked
G$320.4-1.2^\ast$ & $0.17$(Seward \etal \cite{1983ApJ...267..698S}) &
$69$ & $150$[PSR B1509-58] & $63$(Kaspi \etal
\cite{1994ApJ...422L..83K}) & $1.5$ \\
%checked, 
%MSH 15-52, RCW 89
%plerion not seen in radio or optical, only x-ray
SNR$0540-69.3$ & $0.08$(Manchester \etal \cite{1993ApJ...411..756M}) &
$49$ & $50.3$[PSR 0540-69] & $39$(Deeter \etal
\cite{1999ApJ...512..300D}) & $0.8$ \\ % p_0 = 38.7?
%checked, SNR 279.7-31.5 in LMC, SNR 0540-69.3
\hline\hline
\end{tabular}
}
\end{center}
\caption[]{List of composite remnants: $P_0$ is the initial spin
period from the radius method, $P_{\rm t}$ is the current spin period
of systems with identified pulsarsq and $P'_0$ is the calculated
initial spin period by using the braking index.
\label{table:table2}}
\end{table*}

We have collected from the literature $13$ composite supernova
remnants. We list the relevant properties of these systems in Table
\ref{table:table2}. The relative sizes of the plerions are taken from
radio observations, with the exception of G$11.2-0.3$ and G$320.4-1.2$
(marked by asterisks), where only X-ray plerions have been observed.
We apply equation \refnew{eq:includept} to determine the initial spin
period for these systems, adopting $\eta_1 = 1.0$, $\eta_3 = 1.02$,
$E_0 = 1.0\times 10^{51}$ erg, $I = 1.4 \times 10^{45}$ g cm$^2$ (the
value for Crab), and $P_t = 2\pi/\Omega_t = \infty$ in cases where it
is unknown. The same results appear in Figure \ref{fig:result} as
filled triangles. By taking the maximum value for $\eta_3$ we probably
under-estimate the initial spin rate, but likely by no more than a
factor of $2$ judging from Figure \ref{fig:numerical}. In the same
table we also list the initial spin periods ($P_0^\prime$) derived for
$4$ pulsars using braking index measurements.

\section{Discussion}

We have demonstrated a new method for inferring initial spin rates of
pulsars residing in composite supernova remnants.  This method uses
the ratio of the plerion radius and the radius of the supernova shell,
so we dub it as the `radius method'.  Given that typical plerions and
supernova remnants are expected to live for $10^3 - 10^4$ years, we
expect that future observations will increase the population to which
the radius method can be applied.

This method does not require a knowledge of parameters like the
density and magnetic field strength of the ISM, or the distance and
age of the SNR.  However, for the method to apply, the observed PWN
has to be in the subsonic stage of its evolution. If this is not the
case, the method systematically over-estimates the initial spin rate
(see below).  Uncertainties in $E_0$, $\eta_3$ and $\Omega_t$
introduce errors of order unity, as will a deviation from pure
spherical symmetry.  A more important uncertainty of this new method
concerns the influence of synchrotron radiative losses in the plerion,
as parametrized by $\eta_1$.  We tend to under-estimate the initial
spin-rate by ignoring these losses (setting $\eta_{1} = 1$).  We will
illustrate the effect of these uncertainties by considering the $4$
systems in Table \ref{table:table2}, which have known central pulsars,
and independent estimates for their initial periods based on
measurements of the pulsar breaking index.\footnote{G$11.2-0.3$
does not have a braking index measurement. But its inferred initial
spin depends little on this index due to its small age.}

Among these systems, Vela is the oldest and should have entered the
subsonic expansion stage. Its associated pulsar being displaced from
the center of the plerion also testifies in favor of an old age.
%Our method over-estimates Vela's spin-down energy by a factor of $2$
%(or even more if the actual $\eta_3$ at subsonic stage is smaller than
%$1.02$, see Fig. \ref{fig:numerical}) for the stated choice of
%parameters.  It is difficult to pin-down the actual reason for this
%discrepancy.  We give three possible parameter changes which lead to a
%similar initial spin period for Vela as derived by Lyne et al.(1996):
%(1) the supernova explosion could be half as energetic as assumed
%here, (2) the pulsar may be $15$ kyr old, and (3) the braking index is
%as high as $2.4$.
%wyq: change this paragraph according to discussion
Adopting $\eta_3 = 1.02$ and other parameters as stated, the radius
method yields an initial spin-down energy for Vela that is a factor of
$2$ above that inferred by Lyne et al. (\cite{1996Natur.381..497L})
from pulsar braking index measurement. This is surprising as we have
ignored radiative loss. We suggest three possibilities that will
reconcile the two methods: (1) the supernova explosion is half as
energetic as assumed here, (2) the pulsar is $15$ kyr old, and (3) the
braking index is as high as $2.4$ (Lyne \etal
\cite{1996Natur.381..497L} reported $1.4$).  The first option will
raise $P_0$ to $52$ ms while the latter two will reduce $P_0^\prime$
to $41$ ms.
%

%The remaining two systems, G$11.2-0.3$ and G$320.4-1.2$, are young
%($\sim 1.5$ kyr). Their plerions could still be in the supersonic
%expansion stage during which the ratio $R_{\rm pwn}/R_{\rm snr}$
%should exceed that in equation \refnew{eq:PWNra}
%(Fig. \ref{fig:numerical} and \S 2 of van der Swaluw \etal
%\cite{eric}). So we expect the actual value of $\eta_3$ in these
%systems to be of order or higher than $1.02$, the theoretical maximum
%during the subsonic expansion stage.  For example, taking $\eta_3 = 2$
%we find $P_0 = 63$ ms for $G11.2-0.3$. A similar correction may have
%to be applied to G$29.7-0.3$ (Kes 75).
%wyq: changed the above paragraph to
The remaining two systems, G$11.2-0.3$ and G$320.4-1.2$, are young
($\sim 1.5$ kyr). Their plerions could still be in supersonic
expansion during which the value of $\eta_3$ (defined formally through
equation \refnew{eq:introduce} in the supersonic stage) can be higher
than $1.02$, the theoretical maximum during the subsonic expansion
stage (Fig. \ref{fig:numerical} and \S 2 of van der Swaluw \etal
\cite{eric}). For example, taking $\eta_3 = 2$ we find $P_0 = 63$ ms
for $G11.2-0.3$. A similar correction may have to be applied to
G$29.7-0.3$ (Kes 75).
%wyq: end of changes

In contrast to the systems discussed above, SNR $0540-69.3$ is the
only case in which the spin-down energy is significantly
under-estimated for our choice of parameters. The plerion in SNR
$0540-69.3$ is too small if the pulsar was indeed born with an initial
spin of $39$ ms. Manchester \etal (\cite{1993ApJ...411..756M}) have
argued that the radiative loss in this system has reduced the total
energy stored in the plerionic magnetic field and relativistic
electrons to a mere $4\%$ of the integrated spin-down luminosity of
the pulsar. Taking $\eta_1 = 0.04$ we recover a $39$ ms initial
period, with a value of $\eta_3 = 1.3$. This value of $\eta_3$ is
reasonable because SNR $0540-69.3$ is only $760$ years old.  None of
the other $3$ systems discussed here requires strong synchrotron
losses in the plerion, which makes SNR $0540-69.3$ unique. Among the
$4$ pulsars, PSR $0540-69$ is not outstanding either in its surface
magnetic field strength or in its spin-down rate. It is likely born
the fastest, though not by a large margin.

Along the same line, Atoyan (\cite{1999AA...346L..49A}) have argued
that in Crab, strong synchrotron losses in its past can account for
the observed flat spectra of the radio electrons in the plerion. If
this applies to all PWNe, it would significantly change our inferred
initial spin rates. This is illustrated in Fig. \ref{fig:result} for a
strong synchrotron loss of $\eta_1 = 0.05$.
%Currently, the Crab suffers a synchrotron loss of $10\%$ of the
%spin-down luminosity in the X-ray band, the highest in the sample of
%Seward \& Wang (\cite{1988ApJ...332..199S}).

Although the effect of radiative losses on the reliability of our
results remain unclear, we briefly explore the implications of these
results.  Our sample (table \ref{table:table2}) is likely biased
towards fast spinners, which produce plerions that are easier to
detect. Even so, all pulsars in our sample seem to be born spinning
much below the break-up rate. Moreover, their initial spin-periods do
not cluster around $\sim 5-10$ ms, a signature of rapid spin-down by
the gravitationally radiating Rossby-waves which occur immediately
after the supernova collapse (Andersson \etal
\cite{2000ApJ...534L..75A}, Wu \etal \cite{wu2001}). The long initial
periods we find imply an effective angular momentum coupling between
the core and envelope of neutron star progenitors until late into
their evolution. The scattering in the initial periods could either
reflect a different decoupling time in different stars, or a
stochastic process that gives rise to the final angular momentum in
the neutron star, such as an off-centered kick during the collapse
(Spruit \& Phinney \cite{1998Natur.393..139S}).

%\clearpage

We acknowledge discussions with Bram Achterberg, Armen Atoyan and an
anonymous referee.

%\clearpage

%\clearpage
%\vbox{
\end{document}